\documentclass[twocolumn]{aastex63}
\usepackage{natbib}

\graphicspath{{./}{figures/}}

\received{January 20, 2021}
\revised{March 2, 2021}
\accepted{March 12, 2021}
\submitjournal{ApJ}

\shortauthors{Kuroda et al.}

\begin{document}

\title{Implications of High Polarization Degree for the Surface State of Ryugu}

\correspondingauthor{Daisuke Kuroda}
\email{dikuroda@kwasan.kyoto-u.ac.jp}

\author[0000-0002-7363-187X]{Daisuke Kuroda}
\affiliation{Okayama Observatory, Kyoto University, 3037-5 Honjo, Kamogata-cho, Asakuchi, Okayama 719-0232, Japan}
\author[0000-0002-3291-4056]{Jooyeon Geem}
\affiliation{Department of Physics and Astronomy, Seoul National University, Gwanak-gu, Seoul 08826, South Korea}

\author[0000-0001-6156-238X]{Hiroshi Akitaya}
\affiliation{Planetary Exploration Research Center, Chiba Institute of Technology, Tsudanuma, Narashino, Chiba 275-0016, Japan}
\affiliation{Hiroshima Astrophysical Science Center, Hiroshima University, Higashihiroshima, Hiroshima 739-8526, Japan}

\author[0000-0002-0460-7550]{Sunho Jin}
\affiliation{Department of Physics and Astronomy, Seoul National University, Gwanak-gu, Seoul 08826, South Korea}

\author{Jun Takahashi}
\affiliation{Nishi-Harima Astronomical Observatory, Center for Astronomy, University of Hyogo, Sayo, Hyogo 679-5313, Japan}

\author{Koki Takahashi}
\affiliation{Asahikawa Campus, Hokkaido University of Education, Hokumon, Asahikawa, Hokkaido 070-8621, Japan}

\author[0000-0001-9067-7653]{Hiroyuki Naito}
\affiliation{Nayoro Observatory, Nisshin, Nayoro, Hokkaido 096-0066, Japan}

\author{Kana Makino}
\affiliation{Nayoro City University, Nishi 4-jo, Nayoro, Hokkaido 096-8641, Japan}
\author[0000-0003-1726-6158]{Tomohiko Sekiguchi}
\affiliation{Asahikawa Campus, Hokkaido University of Education, Hokumon, Asahikawa, Hokkaido 070-8621, Japan}

\author[0000-0002-2618-1124]{Yoonsoo P. Bach}
\author{Jinguk Seo}
\affiliation{Department of Physics and Astronomy, Seoul National University, Gwanak-gu, Seoul 08826, South Korea}

\author{Shuji Sato}
\affiliation{Astrophysics Department, Nagoya University, Chikusa-ku, Nagoya, Aichi, 464-8602, Japan}

\author{Hiroshi Sasago}
\affiliation{Sasago Co.,Ltd., Chikusa-ku Nagoya, Aichi, 464-0815, Japan}

\author[0000-0001-6099-9539]{Koji S. Kawabata}
\affiliation{Hiroshima Astrophysical Science Center, Hiroshima University, Higashihiroshima, Hiroshima 739-8526, Japan}

\author{Aoi Kawakami}
\author{Miyako Tozuka}
\affiliation{Nishi-Harima Astronomical Observatory, Center for Astronomy, University of Hyogo, Sayo, Hyogo 679-5313, Japan}

\author[0000-0002-3656-4081]{Makoto Watanabe}
\affiliation{Department of Applied Physics, Okayama University of Science, Kita-ku, Okayama, Okayama 700-0005, Japan}
\author[0000-0002-7084-0860]{Seiko Takagi}
\author[0000-0002-6757-8064]{Kiyoshi Kuramoto}
\affiliation{Department of Cosmosciences, Hokkaido University, Kita-ku, Sapporo, Hokkaido 060-0810, Japan}

\author{Makoto Yoshikawa}
\author[0000-0001-6366-2608]{Sunao Hasegawa}
\affiliation{Institute of Space and Astronautical Science, Japan Aerospace Exploration Agency, Sagamihara, Kanagawa 252-5210, Japan}
\author[0000-0002-7332-2479]{Masateru Ishiguro}
\affiliation{Department of Physics and Astronomy, Seoul National University, Gwanak-gu, Seoul 08826, South Korea}



\begin{abstract} 

The asteroid exploration project "Hayabusa2" has successfully returned samples from the asteroid (162173) Ryugu. In this study, we measured the linear polarization degrees of Ryugu using four ground-based telescopes from 2020 September 27 to December 25, covering a wide-phase angle (Sun-target-observer's angle) range from 28$\degr$ to 104$\degr$. We found that the polarization degree of Ryugu reached 53$\%$ around a phase angle of 100$\degr$, the highest value among all asteroids and comets thus far reported. The high polarization degree of Ryugu can be attributed to the scattering properties of its surface layers,  in particular the relatively small contribution of multiply-scattered light. Our polarimetric results indicate that Ryugu's surface is covered with large grains. On the basis of a comparison with polarimetric measurements of pulverized meteorites, we can infer the presence of submillimeter-sized grains on the surface layer of Ryugu. We also conjecture that this size boundary represents the grains that compose the aggregate. It is likely that a very brittle structure has been lost in the recovered samples, although they may hold a record of its evolution. Our data will be invaluable for future experiments aimed at reproducing the surface structure of Ryugu.


\end{abstract}



\section{Introduction} \label{sec:intro}
The Hayabusa2 spacecraft landed on the target C-type asteroid (162173) Ryugu and successfully collected surface materials in its sample-retuning mission. This is the world's first successful attempt to acquire the surface materials from a C-type asteroid that may contain organic compounds and hydrated minerals \citep{Burbine:2008}. This sample was just returned to Earth on  2020 December 6. The new close-up images of Ryugu have raised a new question. Conventionally, it is considered that the surfaces of airless solar system bodies are covered with small dust grains (so-called regolith). Most of these small grains are the products of past impacts with other celestial bodies. It is also known that the temperature difference between day and night generates mechanical stress and breaks up large surface boulders to produce these small dust grains, a process known as thermal fatigue \citep{Delbo:2014}.  

However, from the new high-resolution images, it can be seen that the surface of Ryugu is covered with large lumpy stones without noticeable smaller particles \citep{Sugita:2019, Jaumann:2019}. There are many cobble-sized rocks but no pebble- or sand-sized particles ($\lesssim$1 cm). This fact raises a question as to where the small grains produced by impacts and thermal fatigue have gone. It may be that the surface rocks are covered with small adhesive grains, and  Hayabusa2's cameras could not resolve such small particles because of insufficient resolution, as suggested in \citet{Morota:2020}. Further observational data will be required to follow up on the insufficient resolution of the onboard instruments.


Polarimetry is a powerful tool for investigating the sub-microscopic structures of surficial materials \citep{Geake:1986}. The linear polarization degree $P_\mathrm{r}$ of solar system objects is conventionally defined as $P_\mathrm{r}$ = ($I_\mathrm{N} - I_\mathrm{P}$) / ($I_\mathrm{N} + I_\mathrm{P}$), where $I_\mathrm{N}$ and $I_\mathrm{P}$ are the intensities polarized in directions perpendicular to and parallel to the scattering plane, respectively. $P_\mathrm{r}$ depends on the phase angle (Sun-target-observer's angle, $\alpha$), with a negative branch indicated by low phase angles ($\alpha \lesssim$20$\degr$) and a positive branch at larger phase angles ($\alpha >$20$\degr$) with a maximum value $P_\mathrm{max}$ of approximately $\alpha$ = 100$\degr$. 

So far, there has been no polarimetric measurement of Ryugu. Here we provide the first polarimetric data on Ryugu, measured at four different observatories. Because of observational circumstances during late 2020, our data cover only the positive branch of $P_\mathrm{r}$. For this reason, we mostly focus on the derivation of $P_\mathrm{max}$ in this paper. We introduce our observations and data analysis in Section 2, show our findings in Section 3, and discuss the results in Section 4.

\section{Observations and Data Analysis} \label{sec:obs_ana}
We observed the linear polarization degree of the near-Earth asteroid (NEA) (162173) Ryugu using the polarization modes of four telescopes and their respective mounted instruments: the Multi-Spectral Imager (MSI; \citealt{Watanabe:2012}) using the 1.6-m Pirka telescope at the Hokkaido University Observatory, the Hiroshima Optical and Near-InfraRed camera (HONIR;  \citealt{Akitaya:2014}) using the 1.5-m Kanata telescope at the Higashi-Hiroshima Observatory, the Wide Field Grism Spectrograph 2 (WFGS2; \citealt{Uehara:2004}) using the 2.0-m Nayuta telescope at the Nishi-Harima Astronomical Observatory, and the Triple Range Imager and POLarimeter \#3 (hereafter, TRIPOL;  \citealt{Sato:2019}) using the 1.8-m telescope at the Bohyunsan Optical Astronomy Observatory. Each instrument, mounted at Cassegrain focus, is designed to obtain highly accurate and reliable polarimetric data by rotating a half-wave plate against a fixed polarizer. Detailed information on the observational instruments is summarized in Table \ref{tab:t1}. The HONIR has simultaneous polarization imaging abilities in the visible and near-infrared regions, whereas three visible bands are available for TRIPOL. Only the R$_\mathrm{C}$ band was used in this study.

The methods for processing the data here were similar to those used in our previous studies \citep{Ishiguro:2017, Kuroda:2018, Kuroda:2021}, and we thus omit a detailed description. After instrument-specific image processing (i.e., bias/dark subtraction, flat correction, and cosmic ray removal), we conducted aperture photometry (typically using about 3--4 times the full width at half maximum (FWHM) of the point source as the aperture diameter) of ordinary and extraordinary light on one frame. For TRIPOL without beam-splitting prism, one light source was measured per frame. We noticed that the TRIPOL half-wave plate sometimes did not rotate properly, indicating an unusual signal level of the sky background (because the sky is highly polarized). We examined the background light values and
excluded such data for the following analysis. We calculated the Stokes parameters for a set of four angles of the half-wave plate (0$\degr$, 45$\degr$, 22.5$\degr$, and 67.5$\degr$) and derived the linear polarization degree from these \citep{Tinbergen:1996}. From the multiple sets taken on the same night (i.e., the $N$ columns in Table \ref{tab:t1}), we computed a weighted average. The values regarding the linear polarization degree and position angle of polarization for Ryugu are summarized in Table \ref{tab:t2}, including information on the date of observation, the phase angle, and other parameters. Our data reduction was performed by combining Image Reduction and Analysis Facility (IRAF; \citealt{Tody:1993}) and our own script. The data obtained at each site were corrected for instrumental polarization with unpolarized stars and for the celestial plane with strongly polarized stars \citep{Turnshek:1990, Schmidt:1992}. We also observed the same polarization standard star BD+59d389 \citep{Schmidt:1992} and checked the consistency of the data obtained at all four sites. 


\section{Results} \label{sec:result}
\begin{figure}[h!]
\figurenum{1}\label{fig:f1}
\begin{center}
\includegraphics[width=\hsize]{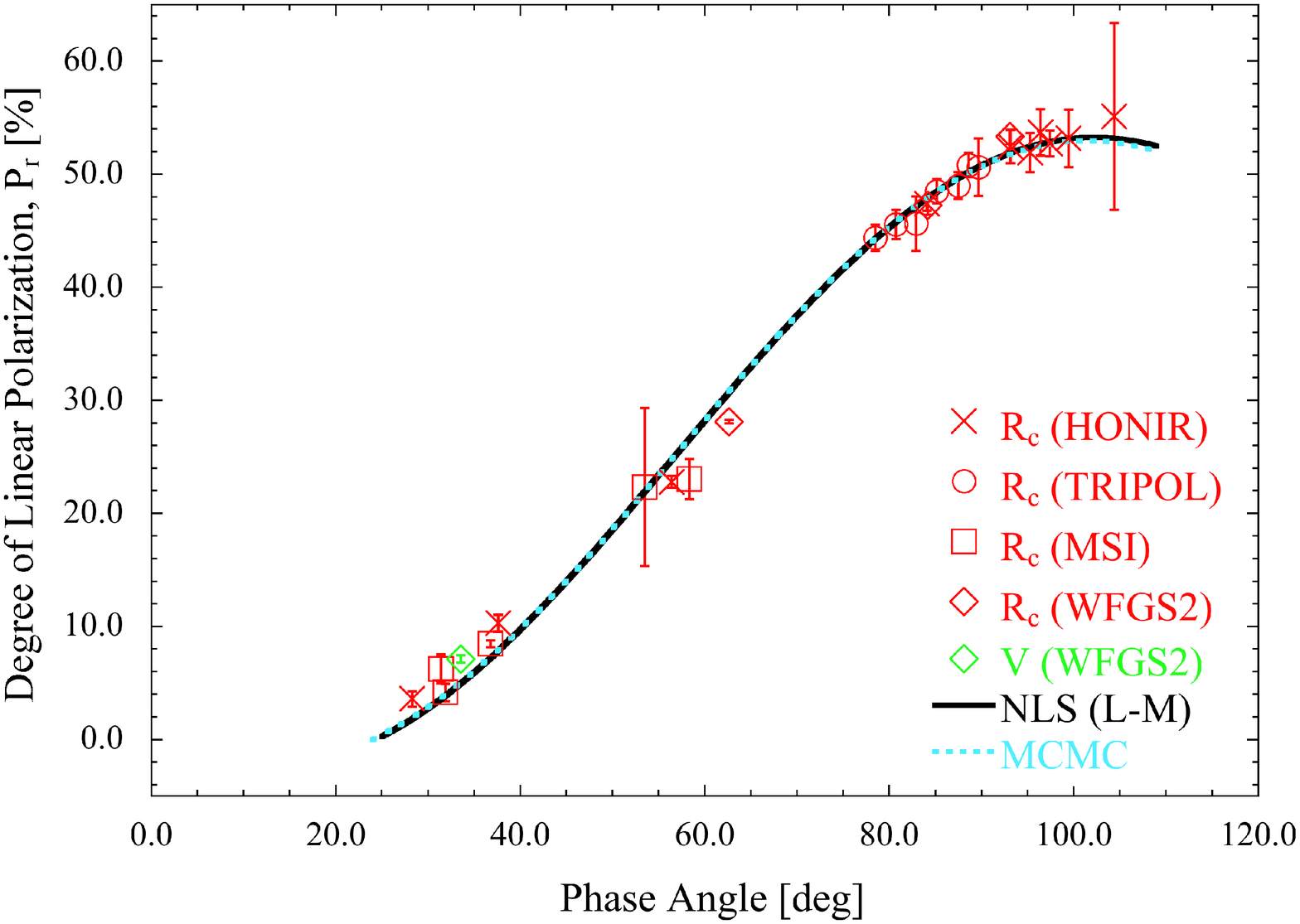}
\caption{Linear polarization degree of Ryugu as a function of the phase angle. The data plotted are the weighted average ($P_r$) of each night shown in Table \ref{tab:t2}. Legend symbols were used separately for each instrument (i.e., crosses for HONIR, open circles for TRIPOL, open squares for MSI, and open diamonds for WFGS2). Red is used for the R$_\mathrm{C}$ band, and green for the V band. The solid black and dotted cyan lines denote the fitting models using NLS method of the L-M algorithm and MCMC method for Equation (\ref{eq:LM}), respectively. In both algorithms, the maximum polarization degree of Ryugu was estimated to be approximately 53\%.}
\end{center}
\end{figure}

\begin{figure}[h!]
\figurenum{2}\label{fig:f2}
\begin{center}
\includegraphics[width=\hsize]{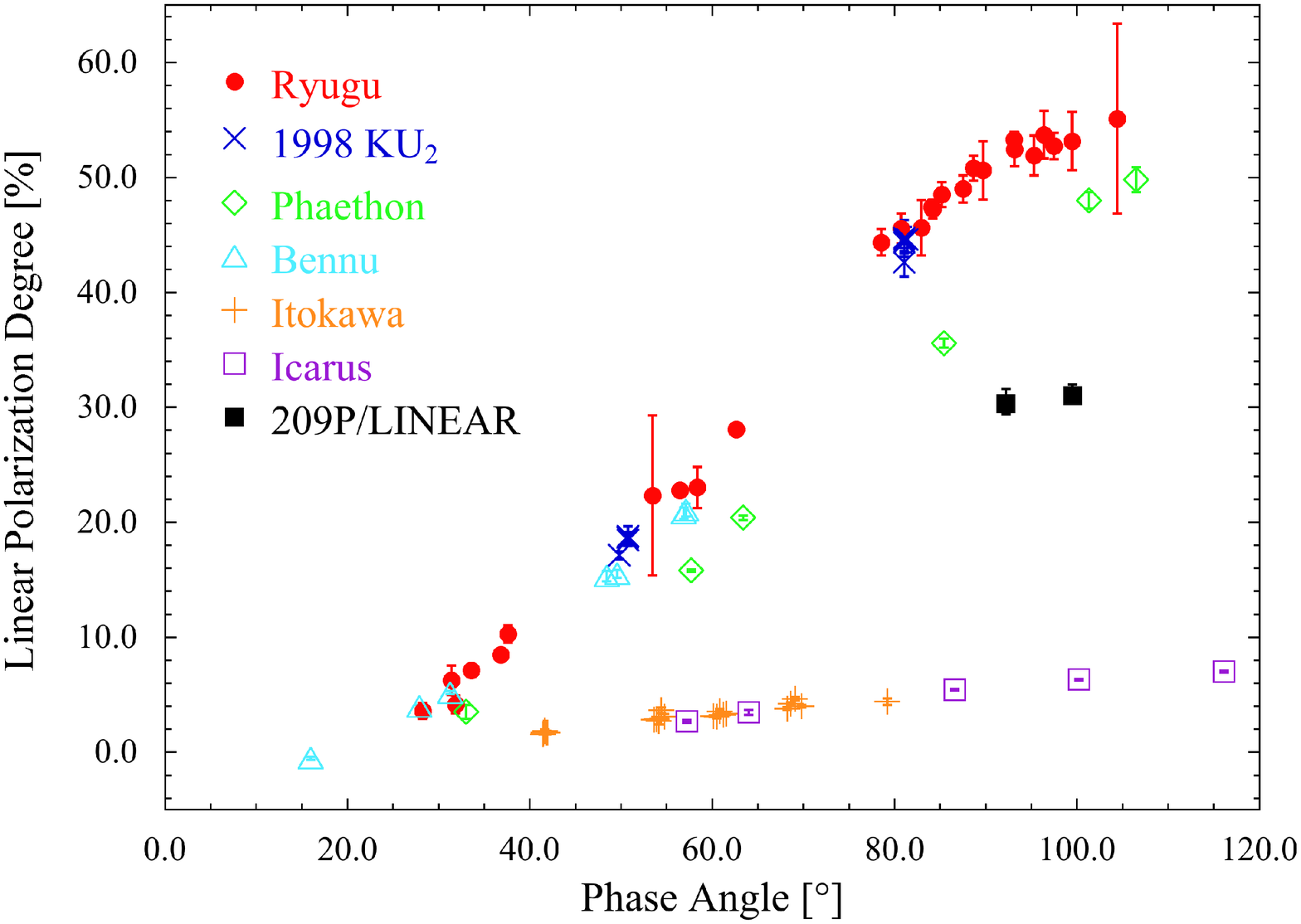}
\caption{Phase angle--polarization dependence of NEAs and cometary nuclei in the red region. The solid red circles represent the polarimetric data for Ryugu. Data for the other NEAs include those from (152679) 1998 KU$_\mathrm{2}$ (blue crosses; \citealt{Kuroda:2018}), (3200) Phaethon (open green diamonds; \citealt{Ito:2018}), (101955) Bennu (open cyan triangles; \citealt{Cellino:2018}), (25143) Itokawa (orange pluses; \citealt{Cellino:2015}), and (1566) Icarus  (open purple squares; \citealt{Ishiguro:2017}). The 209P/LINEAR (solid black squares; \citealt{Kuroda:2018}) correspond to the polarization degree of the cometary nucleus  subtracted its coma component. The linear polarization degree of Ryugu was found to reach the highest value among those measured in existing research.}
\end{center}
\end{figure}

From 2020 September to December, we obtained polarimetric data for Ryugu at 23 nights in the R$_\mathrm{C}$ band and 1 night in the V band, with phase angles ranging from 28$\degr$ to 104$\degr$. Figure \ref{fig:f1} plots the nightly weighted average of the linear polarization degrees for Ryugu versus the phase angle, as well as a corresponding fitting curve. It was found that there was a clear change in the degree of polarization depending on the phase angle. According to our observations, the maximum polarization degree ($P_\mathrm{max}$) of Ryugu reaches $\gtrsim$53\%. Because it was not yet decreasing at this value, the peak may even be a few percent higher than the observed maximum value.
We applied the Lumme and Muinonen function \citep{Lumme:1993, Penttila:2005} for this curve fitting, defined as 
\begin{equation}
P(\alpha) = b\left(\sin \alpha\right)^{c_1} \left( \cos\left(0.5\alpha\right)\right)^{c_2} \sin\left(\alpha-\alpha_0\right), 
 \label{eq:LM}
\end{equation}
\noindent
where $b$, $c_\mathrm{1}$, $c_\mathrm{2}$, and $\alpha_\mathrm{0}$ denote positive constants. 
Although this equation\ref{eq:LM} is an empirical function and may not always simulate the phase angle--polarization degree, the fitting results obtained via both the Markov chain Monte Carlo (MCMC) and nonlinear least squares (NLS) methods (Levenberg-Marquardt (L-M) algorithm) yielded similar values. The expected maximum values were $P_\mathrm{max}$ = $53.0\%^{+1.9\%}_{-1.5\%}$, with the phase angle in this case being $\alpha_\mathrm{max}$ = $101\fdg7^{+8\fdg4}_{-4\fdg8}$ (MCMC), and $P_\mathrm{max}$ = $53.3\%^{+5.9\%}_{-5.1\%}$, with $\alpha_\mathrm{max}$ = $102\fdg2^{+19\fdg1}_{-17\fdg7}$ (L-M). Because no small phase angle data are available for this asteroid, we refrain from mentioning any polarization parameters at low $\alpha$ values (i.e., the minimum value of the polarization degree ($P_\mathrm{min}$),  inversion angle ($\alpha_\mathrm{0}$), and polarimetric slope ($h$)), which is beyond the scope of this work. Figure \ref{fig:f2} compares the phase angle--polarization degree variations in Ryugu, other asteroids, and cometary nucleus. The polarization degree of Ryugu is equal to or even greater than that of (3200) Phaethon \citep{Ito:2018}, which is known to be the highest yet observed. In terms of cometary nucleus, it is also clearly higher than the reported values of $\sim$30\% for 209P/LINEAR \citep{Kuroda:2015}. The phase angle dependence of the polarization degrees for Ryugu is similar to that of (152679) 1998 KU$_\mathrm{2}$ \citep{Kuroda:2018}, whereas it is slightly different from that of  (101955) Bennu \citep{Cellino:2018}, which is the target of the OSIRIS-REx mission. Spectroscopically, Ryugu is classified as a C(Bus-DeMeo\footnote{Bus-DeMeo asteroid taxonomy classification based on visible and near-infrared spectroscopy \citep{DeMeo:2009}})- or Cb(Bus\footnote{Bus asteroid taxonomy classification based on visible spectroscopy \citep{Bus:2002}})-type asteroid \citep{Moskovitz:2013,Tatsumi:2020}, and 1998 KU$_\mathrm{2}$ is also Cb(Bus$^{2}$)-type \citep{Binzel:2004}. The geometric albedo of Ryugu is slightly brighter, at 0.04--0.045 \citep{Sugita:2019,Tatsumi:2020}, than that of 1998 KU$_\mathrm{2}$, at 0.018--0.03 \citep{Mainzer:2011,Nugent:2016}. However, considering the dependability of the absolute magnitude ($H$mag) used to calculate the albedo of 1998 KU$_\mathrm{2}$ \citep{Masiero:2021}, this mismatch is probably within the margin of error. Therefore, we can conclude that Ryugu and 1998 KU$_\mathrm{2}$ have very similar polarimetric and spectroscopic features.

\begin{deluxetable*}{llclcl}
\tablenum{1}
\tablecaption{List of Telescopes and Observational Instruments}
\tablewidth{0pt}
\tablehead{
\colhead{Observatory} & \colhead{Telescope$^{(a)}$} & \colhead{Instrument$^{(b)}$} & \colhead{Camera/Sensor} & \colhead{Pixel Scale} &
\colhead{Polarizer}  \\
\colhead{Oper. Org.$^{(c)}$} & \colhead{Aperture} & \colhead{Reference} & \colhead{Active Pixels} & \colhead{Pixel Size} &
\nocolhead{}
}
\startdata
Hokkaido University & Pirka & MSI & Hamamatsu  C9100-13 & 0.389 \arcsec pixel$^{-1}$     & Wollaston \\
\hspace{15pt}Hokkaido Univ.                             & 1.6 m         &    \citet{Watanabe:2012}    &  512 $\times$ 512       &  16 $\mu$m     \\
Higashi-Hiroshima & Kanata & HONIR & Hamamatsu 2k x 4k$^{(d)}$ & 0.294 \arcsec pixel$^{-1}$ & Wollaston \ \\
\hspace{15pt}Hiroshima Univ.    & 1.5 m          &   \citet{Akitaya:2014}     & 2048 $\times$ 4096      &  15 $\mu$m  \\
Nishi-Harima &  Nayuta & WFGS2 & FLI$^{(e)}$ PL230-42 & 0.198 \arcsec pixel$^{-1}$ & Wollaston  \\
\hspace{15pt}Univ. of Hyogo & 2.0 m          &   \citet{Uehara:2004}     &     2048 $\times$ 2048   &  15 $\mu$m    \\
Bohyunsan &  &  TRIPOL \#3 & SBIG$^{(f)}$ ST-9XEi & 0.29 \arcsec pixel$^{-1}$ & Wire-grid   \\
\hspace{15pt} KASI$^{(g)}$    & 1.8 m          &      \citet{Sato:2019}      & 512 $\times$ 512       &  20 $\mu$m                 \\
\enddata
\tablecomments{
$^{(a)}$Telescope nickname, if any.
$^{(b)}$Abbreviation of the instrument used.
$^{(c)}$Operational Organization.
$^{(d)}$Hamamatsu Photonics fully depleted back-illuminated CCD.
$^{(e)}$Finger Lakes Instrumentation.
$^{(f)}$Santa Barbara Instrument Group.
$^{(g)}$Korea Astronomy and Space Science Institute.
}
\label{tab:t1}
\end{deluxetable*}

\begin{deluxetable*}{cccrrrrrrl} 
\tablenum{2}
\tablecolumns{10} 
\tablewidth{0pc} 
\tablecaption{Summaries of polarimetric results for Ryugu} 
\tablehead{
\colhead{Date} & \colhead{Filter$^{(a)}$}  & \colhead{Exp$^{(b)}$} & \colhead{$N^{(c)}$} & \colhead{$\alpha^{(d)}$} & \colhead{$P \pm \sigma P^{(e)}$}  &  \colhead{$\theta_P \pm \sigma \theta_P^{(f)}$}  & \colhead{$P_r \pm \sigma P_r^{(g)}$}  &  \colhead{$\theta_r \pm \sigma \theta_r^{(h)}$} & \colhead{Inst.$^{(i)}$}\\
\nocolhead{} & \nocolhead{}  & \colhead{(sec)} & \nocolhead{} & \colhead{($\degr$)}  & \colhead{(\%)} & \colhead{($\degr$)} & \colhead{(\%)} & \colhead{($\degr$)} & \nocolhead{}
}
\startdata
2020 Sep 27& R$_\mathrm{C}$ & 120 & 12 & 28.27  & 3.60 $\pm$ 0.68  & 77.90 $\pm$ 5.40 & 3.59 $\pm$ 0.68 & 2.05 $\pm$ 5.40 & HONIR \\ 
2020 Oct 09& R$_\mathrm{C}$ & 60, 120 & 1, 4 &  31.40  & 6.29 $\pm$ 1.29  & -13.13 $\pm$ 4.79 & 6.25 $\pm$ 1.29 & 3.15 $\pm$ 4.79 & MSI \\ 
2020 Oct 10& R$_\mathrm{C}$ & 120 & 3 & 31.89  & 4.19 $\pm$ 0.78  & -19.12 $\pm$ 4.49 & 4.17 $\pm$ 0.78 & -2.84 $\pm$ 4.49 & MSI \\ 
2020 Oct 13& V & 300 & 6 & 33.57 & 7.13 $\pm$ 0.33  & 36.07 $\pm$ 1.32 & 7.12 $\pm$ 0.33 & -1.29 $\pm$ 1.32 & WFGS2\\ 
2020 Oct 18& R$_\mathrm{C}$ & 120 & 12 & 36.79  & 8.53 $\pm$ 0.31  & -19.36 $\pm$ 1.03 & 8.48 $\pm$ 0.31 & -3.08 $\pm$ 1.03 & MSI \\ 
2020 Oct 19& R$_\mathrm{C}$ & 120 & 2 & 37.61  & 10.41 $\pm$ 0.75  & 17.06 $\pm$ 2.07 & 10.29 $\pm$ 0.75 & -4.35 $\pm$ 2.07 & HONIR\\
2020 Nov 06& R$_\mathrm{C}$ & 60, 90 & 1, 1 & 53.48  & 27.65 $\pm$ 6.98  & -34.35 $\pm$ 5.72 & 22.33 $\pm$ 6.98 & -18.07 $\pm$ 5.72 & MSI  \\ 
2020 Nov 09& R$_\mathrm{C}$ & 120 & 12 & 56.43 & 22.82 $\pm$ 0.52  & -10.22 $\pm$ 0.66 & 22.79 $\pm$ 0.52 & -1.60 $\pm$ 0.66 & HONIR \\ 
2020 Nov 11& R$_\mathrm{C}$ & 120 & 3 & 58.33  & 23.10 $\pm$ 1.79  & -18.46 $\pm$ 2.12 & 23.03 $\pm$ 1.79 & -2.18 $\pm$ 2.12 & MSI \\ 
2020 Nov 15& R$_\mathrm{C}$ & 180, 300 & 2, 4 & 62.62  & 28.18 $\pm$ 0.16  & -15.87 $\pm$ 0.22 & 28.11 $\pm$ 0.16 & 2.05 $\pm$ 0.22 & WFGS2 \\ 
2020 Nov 30& R$_\mathrm{C}$ & 60, 120 & 4, 6 & 78.50 & 44.55 $\pm$ 1.16 &  -24.05 $\pm$ 0.74 & 44.37 $\pm$ 1.16 &  -2.53 $\pm$ 0.74 & TRIPOL \\
2020 Dec 02& R$_\mathrm{C}$ & 90 & 22 & 80.71  & 45.69 $\pm$ 1.29 &  -24.01 $\pm$  0.81 &  45.58 $\pm$ 1.29 &  -1.99 $\pm$ 0.81 &  TRIPOL  \\
2020 Dec 04& R$_\mathrm{C}$ & 90 & 5 & 82.94  &  45.79 $\pm$ 2.41 &  -24.73  $\pm$ 1.50 &  45.64 $\pm$  2.41 &  -2.34  $\pm$ 1.50  &  TRIPOL \\
2020 Dec 05& R$_\mathrm{C}$ & 120 & 13 & 84.09  & 47.47 $\pm$ 0.62  & -23.94 $\pm$ 0.38 & 47.41 $\pm$ 0.62 & -1.41 $\pm$ 0.38 & HONIR \\ 
2020 Dec 05& R$_\mathrm{C}$ & 180 & 3 & 84.18  & 47.35 $\pm$ 0.82  & -24.46 $\pm$ 0.45 & 47.24 $\pm$ 0.82 & 1.92 $\pm$ 0.45 & WFGS2 \\ 
2020 Dec 06& R$_\mathrm{C}$ & 90 & 11 & 85.17  & 48.70 $\pm$ 1.07  & -25.20 $\pm$ 0.63 & 48.51 $\pm$ 1.07 & -2.58 $\pm$ 0.63 & TRIPOL\\ 
2020 Dec 08& R$_\mathrm{C}$ & 90 & 22 & 87.49  & 49.02 $\pm$ 1.19  & -23.36 $\pm$ 0.70 & 49.00 $\pm$ 1.19 & -0.66 $\pm$ 0.70 & TRIPOL\\ 
2020 Dec 09& R$_\mathrm{C}$ & 90 & 24 & 88.62  & 50.98 $\pm$ 1.06  & -25.06 $\pm$ 0.60 & 50.81 $\pm$ 1.06 & -2.37 $\pm$ 0.60 & TRIPOL\\ 
2020 Dec 10& R$_\mathrm{C}$ & 90 & 4 & 89.68  & 50.88 $\pm$ 2.54  & -25.55 $\pm$ 1.43 & 50.62 $\pm$ 2.54 & -2.93 $\pm$ 1.43 & TRIPOL\\ 
2020 Dec 13& R$_\mathrm{C}$ & 180 & 6 & 93.10  & 53.41 $\pm$ 0.62  & -23.92 $\pm$ 0.31 & 53.31 $\pm$ 0.62 & 1.75 $\pm$ 0.31 & WFGS2 \\ 
2020 Dec 13& R$_\mathrm{C}$ & 120 & 12 & 93.13  & 52.60 $\pm$ 1.46  & -24.39 $\pm$ 0.80 & 52.44 $\pm$ 1.46 & -2.23 $\pm$ 0.80 & HONIR \\
2020 Dec 15& R$_\mathrm{C}$ & 120 & 9 & 95.29  & 52.28 $\pm$ 1.73  & -25.00 $\pm$ 0.95 & 51.92 $\pm$ 1.73 & -3.57 $\pm$ 0.95 & HONIR \\ 
2020 Dec 16& R$_\mathrm{C}$ & 120 & 3 & 96.40  & 54.12 $\pm$ 2.07  & -24.85 $\pm$ 1.09 & 53.71 $\pm$ 2.07 & -3.83 $\pm$ 1.09 & HONIR \\ 
2020 Dec 17& R$_\mathrm{C}$ & 120 & 11 & 97.44  & 53.08 $\pm$ 1.14  & -24.19 $\pm$ 0.62 & 52.73 $\pm$ 1.14 & -3.29 $\pm$ 0.62 & HONIR \\ 
2020 Dec 19& R$_\mathrm{C}$ & 120 & 5 & 99.47  & 53.62 $\pm$ 2.55  & -23.70 $\pm$ 1.36 & 53.17 $\pm$ 2.55 & -3.73 $\pm$ 2.55 & HONIR \\ 
2020 Dec 25& R$_\mathrm{C}$ & 120 & 2 & 104.38  & 56.14 $\pm$ 8.26  & -21.77 $\pm$ 4.22 & 55.11 $\pm$ 8.26 & -5.49 $\pm$ 4.22 & HONIR \\ 
\enddata 

\tablecomments{
   $^{(a)}$Employed filters.
   $^{(b)}$Typical exposure time per frame in seconds. 
   $^{(c)}$Number of sets counted as one in the four directions of the half-wave plate.
   $^{(d)}$Median phase angle (Sun--Ryugu--Observer angle) in degree. 
   $^{(e)}$Degree of linear polarization in percent.
   $^{(f)}$Position angle of the polarization (in degrees equatorial E of N).
   $^{(g)}$Degree of linear polarization with respect to the scattering plane in percent.
   $^{(h)}$Position angle of the polarization with respect to the normal to the scattering plane (measured anti-clockwise as viewed by the observer, in degrees) 
   $^{(i)}$Abbreviation of the instrument used.
}
\label{tab:t2}
\end{deluxetable*} 

\begin{figure}[h!]
\figurenum{3}\label{fig:f3}
\begin{center}
\includegraphics[width=\hsize]{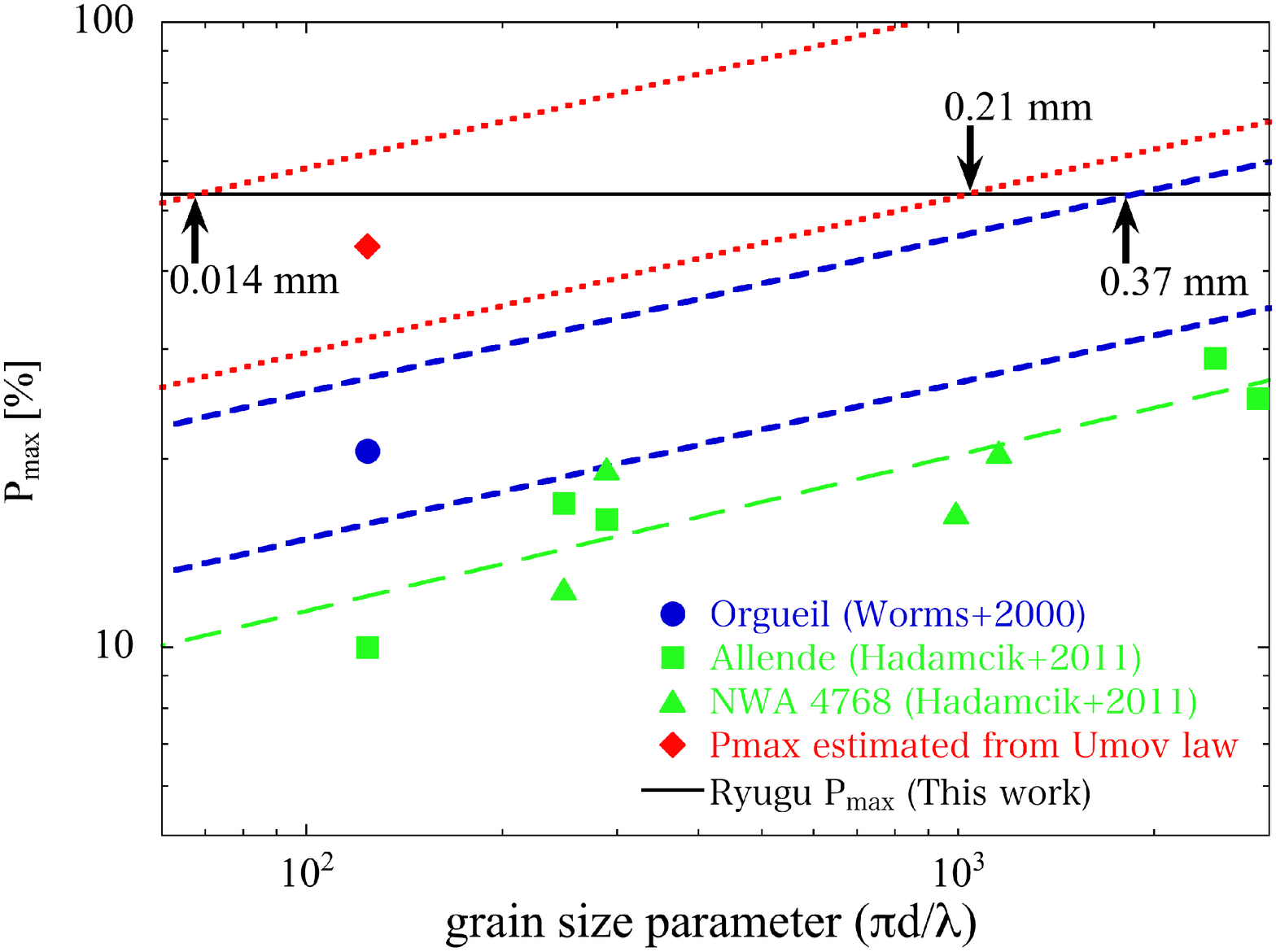}
\caption{Grain size parameter vs. maximum value of the polarization degree ($P_\mathrm{max}$). The size parameter is a dimensionless quantity expressed by $\pi\mathrm{d}/\lambda$, where d and $\lambda$ represent the particle equivalent diameter and wavelength, respectively \citep{Bohren:1998}. The solid green squares and solid green triangles denote two size-sorted carbonaceous chondrites, Allende and NWA 4768 meteorites, respectively \citep{Hadamcik:2011a}. 
The trend of increasing $P_\mathrm{max}$ with grain size for these meteorites is shown by the long green dashed line. Assuming that this proportional relationship including its deviation holds for lower albedo, the range of low albedo ($\sim$0.06) is represented by the blue dashed lines and the range of very low albedo (0.032--0.034, converted from $p_\mathrm{V}$ = 0.04--0.045 for Ryugu) by the red dotted lines. The solid blue circle indicates the pulverized Orgueil meteorite (20--30 $\mu$m), which is the low albedo carbonaceous meteorite \citep{Worms:2000}. The solid red diamond is the reference point of the very low albedo, which has the same grain size of $\sim$25 $\mu$m and is estimated from Umov law \citep{Umow:1905}. From the intersection of this region and Ryugu's $P_\mathrm{max}$ (solid black line), we estimated a grain size of 0.37 mm at the low limit for low albedo ($\sim$0.06) and 0.014--0.21 mm for very low albedo (0.032--0.034).}
\end{center}
\end{figure}
 
\section{Discussion} \label{sec:discussion}
Ryugu has the highest degree of polarization in any air-less solar system object previously observed. Because the polarization degree is determined by the surface conditions, there exists a mechanism that produces a high polarization degree on the surface layer of Ryugu. Umov law \citep{Umow:1905}, which is known as the inverse correlation between albedo and polarization, states that the darker the object, the higher the degree of polarization of the scattered light. The geometric albedo of Ryugu is very low, ranging from 0.04 to 0.045 \citep{Sugita:2019,Tatsumi:2020}. The light-scattering properties of such low albedo are consistent with a qualitatively higher degree of polarization because  it inhibits multiple scattering \citep{Bohren:1998}. Given the polarimetric measurements taken using the returned samples of various shapes and sizes acquired by Hayabusa2, we expect to identify and quantify the physical quantities responsible for the polarization degree.

Although we recognize that the returned sample measurements will narrow down the physical property on Ryugu, as a leading study, we try to constrain the surface state from our polarization data. Generally, multiple scattering is generated by the repeated reflection and refraction of incident light inside an object and is often associated with dense materials. When focusing on the scattering frequency of a surface layer, it is predicted that a larger particle size per unit volume, a lower the scattering frequency. Until now, polarimetric measurements of meteorites (especially with low albedos) have been limited. Among them, the polarization degree of 20- to 30-$\mu$m aggregates of pulverized Orgueil meteorite, the primitive meteorite with low albedo ($\sim$0.06 at $\alpha$ = 6$\degr$, where $\alpha$ means the phase angle), was measured to have a maximum value $P_\mathrm{max}$ = 20.6\% at 638 nm under the deposited condition \citep{Worms:2000}. Polarimetric measurements using slightly brighter meteorites indicated that $P_\mathrm{max}$ was approximately three times higher when the grain size increased from $<$25 $\mu$m to $<$500 $\mu$m for Allende, and approximately 1.3 times higher when the grain size increased from $<$50 $\mu$m to $<$200 $\mu$m for NWA 4768 \citep{Hadamcik:2011a}. That is, as the particle size increases in this size range, the degree of polarization tends to be higher. Applying the relationship between particle size and polarization degree to the lower albedo meteorite (i.e., Orgueil), coupled with the $P_\mathrm{max}$ derived in this study, we obtain the particle size boundary. The bounds of the particle size represent a range that takes into account the error of the applied slope and depends on the data variances of Allende and NWA 4768. We derived a lower limit size of 0.37 mm, but avoided specifying the upper limit, because for some materials, the polarization degree reaches a plateau with increasing size parameter \citep{Hadamcik:2009}, and there were concerns about extrapolation to larger size parameters  (Figure \ref{fig:f3}). This lower limit estimate is consistent with the 0.2- to 2-mm (the square roots of 0.03--4.56 mm$^2$) bright inclusions identified in proximate images of Ryugu \citep{Jaumann:2019}. Considering that the very low albedo of Ryugu ($p_\mathrm{V}$ = 0.04--0.045) is due to the dark matrix, we attempted to derive a polarimetrically dominant grain size for Ryugu. The albedo of Ryugu at $\alpha$ = 6$\degr$, which is adjusted to the measurement conditions of the meteorites, is 0.032 to 0.034 \citep{Tatsumi:2020}. The Ryugu-equivalent reference point (red diamond symbol) shown in Figure \ref{fig:f3} was calculated from Umov law using the  $P_\mathrm{max}$ and albedos of Allende ($\sim$0.11 for $<$25 $\mu$m) and Orgueil ($\sim$0.06 for 20--30 $\mu$m) meteorites. If the above relationship between particle size and polarization degree applies to the very low albedo asteroid (i.e., 0.032--0.034 for Ryugu), we also infer that submillimeter order particles are dominant in the surface layer of Ryugu. In meteorites, the matrix grain size is generally finer than the inclusions, which is qualitatively consistent with our derivation.

On the other hand, the observational results of an optical navigation camera (ONC-T), a radiometer (MARA) and a camera (MasCam) onboard the mobile asteroid surface scout (MASCOT) of Hayabusa2 reported that the surface layer of Ryugu was covered with submeter- to meter-sized rocks, with no fine-grain dust below the millimeter order \citep{Sugita:2019,Grott:2019,Jaumann:2019}. On the basis of the result of a thermal infrared imager (TIR), the rocks of the Ryugu surface are much more porous than typical carbonaceous chondrite meteorites, and the surrounding soils are similarly porous and covered with small pieces of rock $>$10 cm in diameter \citep{Okada:2020}. By regarding it as a thermally consolidated block, the size of the domain particles that comprise the rock is not constrained. Data from thermal infrared observations indicate that the surface layer of Ryugu has a very high porosity \citep{Grott:2019,Ogawa:2019,Shimaki:2020}. Thus we also consider that our estimate (i.e., submillimeter-sized grains) corresponds to particle sizes that constitute highly porous aggregates, whose agglomerates form centimeter to submeter rocks. A possible interpretation of the size (determined by our polarimetry) is that the cauliflower-like structures on the rock appearances observed by MASCOT are attributed to determining the polarimetric property. 

Although we acknowledge that our size estimate is crude because of the unavailability of polarimetric measurements of very dark meteorites, it seems that very small (micron-sized) grains are not so abundant as to reduce the $P_\mathrm{max}$. It is likely that the paucity of micron-sized particles may have been selectively ejected by some mechanisms such as levitation \citep{Hartzell:2019}, or may have been buried underground \citep{Arakawa:2020} or have been incorporated into larger grains during aggregate formation \citep{Seiphoori:2020}. The most important contribution of this research is to provide basic polarimetric data for reproducing the physical conditions of the Ryugu surface layer. Further polarimetric investigation using Ryugu samples will specify the surface physical conditions. 

\section{Summary} \label{sec:summary}
We conducted polarimetric observations of the NEA (162173) Ryugu, which was thoroughly investigated by the Hayabusa2 spacecraft, from 2020 September to December. The degree of linear polarization of Ryugu was obtained for 24 nights over phase angles of 28$\degr$--104$\degr$, and the following findings were obtained.
 \begin{enumerate}
\item The polarization degree of Ryugu, which reached $\gtrsim$53\%, is the highest ever recorded, exceeding that of Phaethon.
\item Submillimeter order grains are dominant in the surface layer of Ryugu, as inferred from comparison with polarimetric measurements of meteorites.
\item The results of in-situ observations (MARA, MasCam, ONC-T, and TIR) of Hayabusa2, also indicate that this grain size reflects the constituents of larger aggregates.
\end{enumerate}
This study provides basic data that allows tracing of the brittle and fragile surface layers of Ryugu in the future.

\acknowledgments

MI was supported by the NRF funded by the Korean Government (MEST) grant No. 2018R1D1A1A09084105 and the Seoul National University Research Grant in 2018. The work of SH was supported by JSPS KAKENHI (grant nos. JP18K03723, JP19H00719, and JP20K04055) and the Hypervelocity Impact Facility (former facility name: the Space Plasma Laboratory), ISAS, JAXA. 
SJ was supported by the Global Ph.D. Fellowship Program through a National Research Foundation of Korea (NRF) grant funded by the Korean Government (NRF-2019H1A2A1074796). 
DK was funded by the Kyoto University Foundation.
This research was partially supported by the Optical \& Near-Infrared Astronomy Inter-University Cooperation Program, MEXT, of Japan. Our observation at the Bohyunsan Optical Astronomy Observatory was realized with the help of the observatory staff, especially Dr. Hyun-il Sung. We would like to thank the staff of the Nayoro Observatory for their cooperation in organizing our observation.

%

\facilities{Kanata:1.5m, Nayuta:2.0m, Pirka:1.6m, BAAO:1.8m}


\software{astropy \citep{Astropy:2013},  IRAF \citep{Tody:1993}, SExtractor \citep{Bertin:1996}, SEP \citep{Barbary:2016}, PyMC3 \citep{Salvatier:2016}}



\bibliography{Kuroda2021}



\end{document}